\documentclass[]{JFM-FLM_Au}

\usepackage{mathtools, amsfonts, amssymb, graphicx,xcolor,tikz,tkz-euclide, booktabs, tensor, empheq,subcaption,lipsum,float,multicol}
\usepackage[capitalize]{cleveref}
\usepackage{siunitx}
\usepackage[T1]{fontenc}
\crefname{equation}{}{}

\newcommand{\ovl}[1]{{\overline{#1}}}

\newcommand{\dave}[1]{#1}

\newcommand{\nRe}{\mathit{Re}}

\newcommand{\nBo}{\mathit{Bo}}

\lefttitle{D. Darrow, L. Warwaruk, \& J.W.M. Bush}

\title{Capillary currents and viscous droplet spreading}

\author{David Darrow\aff{1}, Lucas Warwaruk\aff{2}, \and John W.~M.~Bush\aff{1}}

\affiliation{\aff{1}Department of Mathematics, Massachusetts Institute of Technology, Cambridge, MA 02139, USA
\aff{2}Department of Mechanical Engineering, Massachusetts Institute of Technology, Cambridge, MA 02139, USA}

\corresau{David Darrow, \email{ddarrow@mit.edu}}

\begin{document}
\maketitle

\begin{abstract}
\dave{We present the results of a combined experimental and theoretical study of the spreading of viscous droplets over rigid substrates}. First, we experimentally investigate the wetting of a roughened glass surface by a viscous droplet of silicone oil, wide and shallow relative to the capillary length $\ell_c$. The \dave{horizontal} radius of the droplet grows according to an $R_\mathrm{drop}\sim t^{1/8}$ scaling reminiscent of \dave{viscous} gravity currents \citep{Lopez1976}. The droplet is preceded by a mesoscopic fluid film that percolates through the rough substrate, its radius increasing according to $R_\mathrm{film}\sim t^{3/8}/(\log t)^{1/2}$. To rationalize these observed scalings, we develop a new `capillary current' model for the spreading of shallow droplets with arbitrary radius on \dave{rough surfaces. Furthermore, on the basis of established similarities between droplet spreading over wetted rough and smooth substrates \citep{Cazabat1986}, we argue its relevance to a broader class of spreading problems}. We propose that, throughout their evolution, \dave{shallow droplets} maintain a quasi-equilibrium balance between hydrostatic and curvature pressure, perturbed only by unbalanced contact line forces arising along the droplet's edge. \dave{For} drops \dave{with horizontal radii} small with respect to $\ell_c$, \dave{our model converges to the original description of \citet{Hervet1984} and thereby recovers the classic spreading laws of \citet{Hoffman1975}, \citet{Voinov1976}, and \citet{Tanner1979}}. For drops wide with respect to $\ell_c$, it \dave{rationalizes why} millimetric, surface-tension-driven \emph{capillary currents} exhibit the same spreading behavior as relatively large-scale viscous gravity currents. 
\end{abstract}



\section{Introduction}\label{sec:intro}

A volume of liquid placed onto a solid substrate will spread in response to gravitational and interfacial forces, with its spreading resisted by a combination of inertial and viscous stresses. The relative importance of these four effects \dave{is prescribed by} the \emph{Bond number} $\nBo = \rho gh^2/\sigma$ and the \emph{Reynolds number} $\nRe = \rho Uh/\mu$. Here, $\sigma$ is the surface tension at the air-liquid interface, $\rho$ and $\mu$ are the liquid's density and dynamic viscosity, $h$ is its \dave{maximum} depth, $U$ is its characteristic speed, and $g$ is the acceleration due to gravity. \dave{While our work primarily focuses on the spreading of viscous, shallow droplets ($\nRe\ll 1$, $\nBo\lesssim 1$), it is helpful to first discuss the relatively well-established theory of {viscous gravity currents} ($\nRe\ll 1$, $\nBo\gg 1$).}

\dave{If $\nRe\ll1$ and $\nBo\gg 1$, corresponding to a deep, viscous volume of liquid, then the fluid spreads out in the form of a \emph{viscous gravity current}~\citep{Simpson1999,Huppert2006,Ungarish2009}}. Suppose it has a horizontal radius $R$ and characteristic volume $V\sim hR^2$. Gravity creates an overpressure $\rho g h$ in the center of the volume, and thus a horizontal pressure gradient $\rho g h/R$ that drives the fluid outward. This force is balanced by a viscous stress $\mu\dot{R}/h^2$, yielding the classic \dave{scaling}~\citep{Huppert1982}
\begin{equation}\label{eq:huppert}
    \dot{R}\sim \frac{\rho g}{\mu}\,\frac{h^3}{R}\sim \frac{\rho g}{\mu}\,\frac{V^3}{R^7},\qquad R\sim V^{3/8}t^{1/8}.
\end{equation}
\dave{This gravity-viscosity balance is known to drive a number of macroscopic flows in nature and industry, including creeping lava flows, molten glass flows in the production of sheet glass, and honey spreading over toast~\citep{Huppert2006}.}

\dave{ The scaling \cref{eq:huppert} holds until the depth $h$ becomes comparable to the capillary length $\ell_c=(\sigma/\rho g)^{1/2}$, at which point one expects interfacial effects to become important. However, the scaling $R\sim V^{3/8}t^{1/8}$ does not change when a \emph{wide} droplet enters this latter regime, specifically when $h\lesssim \ell_c$ and $R\gg \ell_c$ \citep{Bonn2009}. \citet{Lopez1976} proposed that wide, shallow droplets behave similarly to deeper gravity currents---namely, that spreading is driven by gravitational overpressure and resisted by energy dissipation throughout the droplet bulk. Subsequent work of \citet{Ehrhard1991} offered an alternate, edge-localized form of the energy dissipation, and \citet{Voinov1995, Voinov1999} suggested perturbatively incorporating the effect of surface tension on the droplet edge. Even still, the basic physical picture of gravity-driven motion in this small-Bond-number regime remains widely accepted \citep{Bonn2009,Popescu2012}}. 

\dave{In the present work, we propose an alternative perspective on the spreading of wide drops. First, we experimentally investigate the spreading of a wide, shallow drop of silicone oil over a roughened glass surface. Silicone oil is totally wetting with respect to borosilicate glass, so the drop exhibits a \emph{wet Cassie state}~\citep{deGennes2003}, in which a thin film of liquid percolates through the rough substrate ahead of the drop's advancing edge. We refer to such a film as a `Darcy precursor film' in order to distinguish it from the microscopic precursor films that appear on smooth substrates~\citep{Hardy1919}. As first established by \citet{Cazabat1986}, the similarities between Darcy and microscopic films serve to create strong dynamical similarities between the total wetting of rough and smooth substrates. In our case, we observe that the horizontal droplet radius $R_1$ grows according to $R_1\sim t^{1/8}$, consistent with existing literature, but the radius $R_2$ of the thin film closely adheres to a \dave{new} scaling $R_2\sim t^{3/8}/(\log t)^{1/2}$.}

To rationalize these findings, we develop a new \dave{`capillary current'} model for the spreading of shallow, viscous droplets, with \dave{horizontal} radius either small or large with respect to $\ell_c$, over either rough or smooth substrates. \dave{In short, we propose that such a droplet maintains a quasi-equilibrium balance between hydrostatic and curvature pressure, perturbed only by unbalanced contact line forces along its edge}. Our proposed model \dave{converges to} the original description of \cite{Hervet1984} in the small-droplet limit, \dave{thereby recovering the classic scalings of \citet{Hoffman1975}, \citet{Voinov1976}, and \citet{Tanner1979}. Namely, in the small-droplet limit ($R_1\ll \ell_c$), the droplet retains the shape of a spherical cap, and the droplet radius $R_1$, volume $V$, and apparent contact angle $\theta$ satisfy
\[\dot{R}_1\sim \theta^3, \qquad R_1\sim V^{3/10}t^{1/10}.\]
In the large droplet limit ($R_1\gg\ell_c$), our model offers a self-consistent rationale for the $R_1\sim V^{3/8}t^{1/8}$ scaling first observed by \citet{Lopez1976}, and clarifies the nature of the small-to-large-droplet transition observed by \citet{Cazabat1986} and \citet{Ehrhard1993}. In addition to our own experimental results for large drops on rough substrates, our model rationalizes the data of \citet{Dorbolo2021} for the spreading of Darcy precursor films ahead of small droplets, and the data of \citet{Cazabat1986} for the small-to-large-droplet transition over smooth substrates. Our model also extends naturally to the case of \emph{partial wetting}, where it reproduces the predictions of \citet{deRuijter1999} and \citet{Durian2022} for the early-time spreading behavior of partially wetting drops. Taken together, our study suggests commonalities between large and small drops spreading over both rough and smooth substrates, under conditions of both total and partial wetting.} 

\dave{We present the basic physical picture of viscous droplet spreading in \cref{sec:physpic}, and provide a brief review of relevant literature. In \cref{sec:experiment}, we present our own experimental results on the spreading of large drops of silicone oil over roughened glass. We introduce our theoretical model in \cref{sec:model} to rationalize the observed spreading behavior. In \cref{sec:priordata}, we discuss the broader relevance of our model, which suggests a new perspective on the experiments of \citet{Cazabat1986}, \citet{deRuijter1999}, and \citet{Dorbolo2021}. We propose additional experiments and directions of further study in \cref{sec:discussion}.}


\section{Physical Picture}\label{sec:physpic}
When a liquid droplet is placed on a rigid substrate, its evolution depends on several \dave{geometric and chemical factors. Of central importance are (a) whether spreading is resisted by viscosity or inertia, (b) whether the drop is shallow, deep, narrow, or wide with respect to the capillary length $\ell_c=(\sigma/\rho g)^{1/2}$, (c) whether the drop \emph{totally} or \emph{partially} wets the substrate, and (d) whether the substrate itself is rough or smooth. We focus here on the case of shallow, viscous drops:
\[\nBo = \rho gh^2/\sigma\lesssim 1,\qquad \nRe = \rho Uh/\mu \ll 1.\]
Within this regime, we consider both `small' ($R\ll\ell_c$) and `large' ($R\gg\ell_c$) drops, and both smooth and rough substrates. In the latter case, we assume the roughness scale $h_2$ of the substrate is small relative to the droplet depth: $h_2\ll h\lesssim \ell_c$. }

\dave{For one, the extent to which the liquid wets the substrate is strongly influenced by surface chemistry. Indeed, if $\gamma_\mathrm{sl}$ and $\gamma_\mathrm{sv}$ are the solid-liquid and solid-vapor interfacial tensions, respectively, the regimes of \emph{total wetting} and \emph{partial wetting} can be distinguished by the spreading parameter
\begin{equation}\label{eq:spreadingparam}
    S = \gamma_\mathrm{sv} - \gamma_\mathrm{sl} - \sigma.
\end{equation}
If $S<0$, then the droplet \emph{partially wets} the substrate, and will spread until reaching a static equilibrium marked by an equilibrium contact angle $\theta_\mathrm{eq}>0$. In this case, the horizontal force balance at the contact line yields Young's Law:
\begin{equation}\label{eq:eq_contactangle}
    \sigma\cos\theta_\mathrm{eq} + \gamma_\mathrm{sl} = \gamma_\mathrm{sv}\qquad \implies \qquad \cos\theta_\mathrm{eq} = 1+ S/\sigma,
\end{equation}
which ensures $\theta_\mathrm{eq}>0$ provided that $S<0$. Alternatively, if $S>0$, then no such balance is possible, and the droplet \emph{totally wets} the substrate: the drop spreads until its depth becomes comparable to the roughness scale of a rough substrate, or the molecular scale for a smooth substrate. We note that this physical picture can be further complicated if, for example, the liquid is volatile or the substrate has deep, porous structure. While such issues are beyond the scope of our study, we direct the reader to \citet{Bonn2009}, \cite{Popescu2012}, and \citet{Gambaryan2014} for reviews of a broader class of spreading problems.}

\begin{figure}
    \centering
    \includegraphics[width=\linewidth]{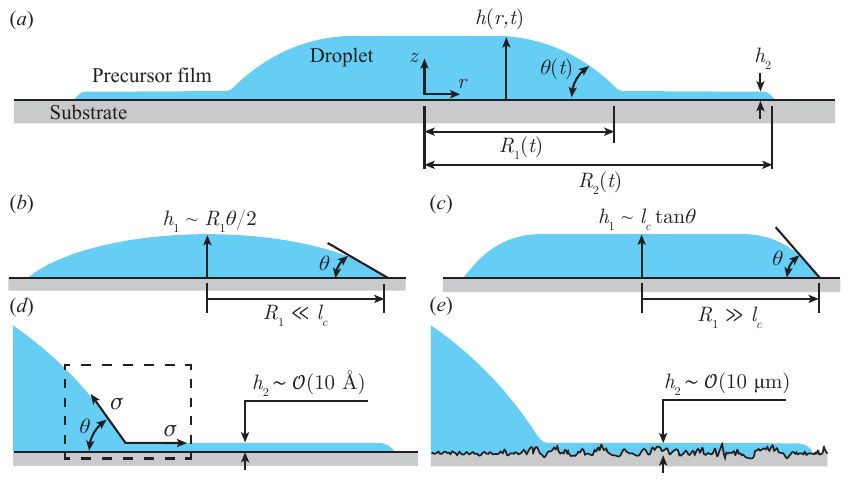}
    \caption{Schematic of the physical system under investigation. \textbf{(a)} The droplet forms the rounded shape \cref{eq:drop_profile} of radius $R_1$, contact angle $\theta$, and \dave{maximum} depth $h_1=h(0)\ll\ell_c$. \dave{In the case of total wetting}, the droplet \dave{is generally} preceded by a precursor film of \dave{thickness} $h_2\ll h_1$ and radius $R_2\geq R_1$. \textbf{(b)} A small droplet ($R_1\ll\ell_c$) approximately forms a spherical cap with a maximum \dave{depth} $h_1 \sim R_1\theta/2$. \textbf{(c)} A large droplet ($R_1\gg\ell_c$) is shaped more like a flat disk, with a \dave{depth} $h_1 \sim \ell_c \tan\theta$ that is approximately constant for $r < R_1 - \ell_c$. \dave{\textbf{(d)} If the substrate is smooth, a \emph{microscopic} precursor film arises with thickness $h_2$ on the \dave{molecular scale} \textbf{(e)} If the substrate is rough, a \emph{Darcy} precursor film arises with thickness comparable to the roughness height (in the case of our experiments, about 10 \SI{}{\micro\meter})}. The zoom box in \textbf{(d)} illustrates the net horizontal surface tension force $f=\sigma(1-\cos\theta)\sim\sigma\theta^2/2$ acting radially outward on the apparent contact line; such a force is present on both smooth and rough substrates. In all cases, the radius $R_1$ grows with time, while $\theta$ and $h_1$ decrease. }
    \label{fig:diagram}
\end{figure}

\dave{We primarily focus on the case of \emph{total wetting}, for which $S>0$, though we return to the question of partial wetting in \cref{sec:priordata}. A drop totally wetting a rough substrate gives rise to a `Darcy precursor film' of liquid that percolates ahead of the drop's advancing edge~\citep{deGennes2003}, with depth comparable to the roughness scale $h_2$ (\cref{fig:diagram}(\textit{e})). Notably, a similar physical picture holds for smooth substrates: it was discovered by \citet{Hardy1919} that the total wetting of a smooth substrate is accompanied by the formation of a \emph{microscopic} precursor film that extends ahead of the visible droplet (\cref{fig:diagram}(\textit{d})). In both cases, the drop spreads over a liquid layer, never coming in direct contact with the solid. The presence of the precursor film on a smooth substrate provides a resolution to the classic paradox of contact line motion~\citep{Huh1971,deGennes1985,Joanny1986}.}

\dave{We discuss the leading-order geometry of the droplet in \cref{sec:geometry}, review the existing models for viscous droplet spreading in \cref{sec:existingresults_small,sec:existingresults_large,sec:existingresults_partial}, and review the existing model for the spreading of Darcy precursor films in \cref{sec:washburn}}. A theoretical model that revises both physical pictures (i.e., for both droplet and film spreading) will be introduced in \cref{sec:model}. \dave{We do not treat the spreading of microscopic precursor films in the present work, but refer the reader to the reviews of \citet{Bonn2009} and \cite{Popescu2012} for a detailed discussion of the topic.} 

\subsection{Quasi-equilibrium Geometry}\label{sec:geometry}

\dave{The basic geometry of our system is depicted in \cref{fig:diagram}. We suppose for simplicity that the apparent contact angle $\theta$ satisfies $\theta\lesssim\pi/4$ (so that $\sin\theta\approx\tan\theta\approx\theta$) and that the droplet is axisymmetric. Here and below, we fix non-dimensionalized units such that
\begin{equation}\label{eq:units}
    \frac{\sigma}{\rho g} = \frac{\sigma}{\mu} = 1,\qquad p_\mathrm{atm.} = 0.
\end{equation}
Equivalently, we non-dimensionalize by the length, velocity, and time scales
\[
    \ell_c = \sqrt{\sigma/\rho g},\qquad u_c = \sigma/\mu,\qquad \tau_c = \ell_c/u_c,
\]
and we tare the pressure such that $p_\mathrm{atm.}=0$.} 

To a good approximation, \dave{after a momentary equilibration period when the droplet is first placed on a substrate \citep{Biance2004,Eddi2013}, the instantaneous droplet shape is determined by a balance of hydrostatic and curvature pressures. Such a balance is well understood to underlie the leading-order droplet geometry, and much of the following argument has appeared, for instance, in the work of \citet{Tanner1979} and \citet{Voinov1999}.}

\dave{We} write $r$ for the radial coordinate and $h(r)$ for the height of the droplet's surface. To leading order in \dave{$h/\ell_c\lesssim 1$}, the pressure immediately inside the droplet satisfies $p(r) = -\sigma \nabla^2 h$. \dave{Balancing hydrostatic and curvature pressures of the surface requires that}
\[\sigma\partial_z\nabla^2 h = \sigma \frac{\partial_r\nabla^2 h}{\partial_r h} = \rho g.\]
\dave{By nondimensionalizing and enforcing axisymmetry, one may find the solution}
\begin{equation}\label{eq:drop_profile}
    h(r) = \frac{I_0(R_1) - I_0(r)}{I_1(R_1)}\,\tan\theta,
\end{equation}
\dave{which has an associated total droplet volume}
\begin{equation}\label{eq:volume}
    V = \int_0^{R_1} 2\pi r h(r)\,dr = \left(\pi R_1^2\frac{I_0(R_1)}{I_1(R_1)} - 2\pi R_1\right)\tan\theta.
\end{equation}
\dave{Here, $R_1=R_1/\ell_c$ is the dimensionless horizontal droplet radius, $\theta$ is the apparent contact angle at the droplet-film interface, and $I_0$ and $I_1$ are modified Bessel functions. As one might expect, this height profile always describes a rounded shape, positive only for $r<R_1$. Its qualitative behavior is best understood by investigating the small- and large-droplet limits. The Bessel functions $I_0$ and $I_1$} have the following asymptotic behavior for small and large arguments:
\begin{equation}\label{eq:asymptotic}
    \begin{aligned}
        x\ll 1\quad&\implies\quad I_0(x)\sim 1 + x^2/4,\quad I_1(x)\sim x/2,\hspace{2em}\\
        x\gg 1\quad&\implies\quad I_0(x)\sim I_1(x)\sim e^x/\sqrt{2\pi x}.
    \end{aligned}
\end{equation}
As \dave{evident} from these asymptotic expressions, the qualitative shape of the droplet \dave{greatly} depends on \dave{the relative magnitudes of $R_1$ and $\ell_c$}. When $R_1\ll \ell_c$ and $\theta\lesssim\pi/4$, the profile \cref{eq:drop_profile} asymptotes to that of a spherical cap with \dave{radius of curvature} $R_\mathrm{cap}\sim R_1/\theta$ and maximum height $h_1\sim R_1\theta/2$, \dave{as} illustrated in \cref{fig:diagram}(\textit{b}). When $R_1\gg \ell_c$, the profile asymptotes to that of a flat puddle of height $h_1\sim \ell_c\tan\theta$, except in the boundary region $|R_1-r|\lesssim \ell_c$, where it decays to zero (\cref{fig:diagram}(\textit{c})). 

\dave{The profile \cref{eq:drop_profile} interpolates smoothly between these two limits.  In particular, if the radius of an initially-small droplet grows to exceed $\ell_c$, the droplet transitions through a sequence of increasingly-flattened spherical shapes until reaching the pancake-like form depicted in \cref{fig:diagram}(\textit{c}) \citep{Cazabat1986,Ehrhard1993}. The precise crossover radius is on the order of $\ell_c$, but has been reported to depend weakly on the droplet volume~\citep{Cazabat1986}.}


\subsection{Existing Models for Total Wetting by Small Droplets}\label{sec:existingresults_small}
\dave{The modern understanding of small droplet spreading ($R_1\ll\ell_c$) over a flat substrate began with the experimental work of \citet{Hoffman1975}, who observed the scaling $\dot{R}\sim \theta^3$. An initial attempt to rationalize Hoffman's findings was provided by \citet{Voinov1976} and \citet{Tanner1979}, based on a lubrication approximation of the droplet dynamics. Their work rationalized \emph{Tanner's laws} (sometimes known as the Hoffman--Voinov--Tanner laws) for the spreading of a small droplet on a rough substrate: $R\sim V^{3/10}t^{1/10}$ and $\theta\sim V^{1/10}t^{-3/10}$. A more complete description of the droplet motion---and the first to correctly account for the precursor film---was subsequently provided by \citet{Hervet1984}. They argue that small droplets maintain the shape of a spherical cap \dave{(see \cref{fig:diagram}(\textit{b}))} as they are slowly stretched outward by contact line forces. In turn, the rate of spreading can be deduced from the droplet's internal energy balance}.

\dave{The argument of Hervet \& de Gennes proceeds as follows.} Suppose the droplet boundary advances at a speed $U$. On one hand, there is a force $f=\sigma(1-\cos\theta)\approx \sigma\theta^2/2>0$ per unit arclength along the apparent contact line---arising from unbalanced surface tension forces when the droplet meets the surface at an angle $\theta\lesssim\pi/4$ (see \cref{fig:diagram}(\textit{d}))---which does work $w\sim \sigma\theta^2 U/2$ per unit arclength. \dave{Hervet \& de Gennes argue that this work is dissipated primarily along the outer edge of the droplet. To model this process, they define a radial coordinate $\chi = R_1-r$ measuring the distance (inward) from the edge}. In order to ensure that the droplet surface (at $z = \theta\chi$) travels with velocity $U$, a linear shear profile $u(\chi,z)\sim Uz/\theta \chi$ \dave{is} adopted. The resulting energy dissipation rate is not integrable, so \dave{the authors} fix upper and lower cutoff lengths in $\chi$, say, $L$ and $a$, respectively. Physically, one expects $L\lesssim\ell_c$ and for $a$ to be on the molecular scale; \dave{indeed, one generally needs a microscopic cutoff length in order to regularize dissipation near a moving contact line \citep{Huh1971}}. \dave{The energy dissipation per unit arclength is thus}
\[\int_{a}^L\int_0^{\theta \chi} \mu\left(\frac{\partial u}{\partial z}\right)^2\,dz\,d\chi = \int_{\ell_0}^L\frac{\mu U^2}{\theta \chi}\,d\chi = \frac{\mu}{\theta}\log(L/a)U^2.\]
Writing $\ell_D=\log(L/a)$, \dave{they thus deduce} the evolution equation $\dot{R}_1\sim U \sim \theta^3/2\ell_D$, consistent with \dave{the predictions of \citet{Hoffman1975}, \citet{Voinov1976}, and \citet{Tanner1979}.}

\dave{Hervet \& de Gennes went a step further and estimated the cutoff lengths $L$ and $a$ based on the micro-structure of the droplet-film transition on a smooth substrate. They thus find a logarithmic correction to Hoffman's scaling, of the form
\begin{equation}\label{eq:logcorrection}
    \theta^3 \sim \dot{R}_1\log \left([\dot{R}_1]^{2/3}\right).
\end{equation}
The work of \citet{Eggers2004} has since offered a slight correction to the scaling coefficients predicted by Hervet \& de Gennes, and separate work by \citet{Hocking1982}, \citet{Hocking1983}, \citet{Cox1986,Cox1986b}, \citet{Snoeijer2006}, \citet{Chan2013}, and \citet{Luo2025} has clarified the local structure of a moving contact line. The physical model furnished by the spreading laws of Hervet \& de Gennes and subsequent work on the contact line problem remains well-supported by a variety of experimental investigations of liquid spreading on smooth substrates \citep{Ngan1982,Chen1988,Chen1989, DussanV1991,Marsh1993,Rame1996}}. 

\dave{The work of \citet{Cazabat1986} showed that the same spreading laws apply to the spreading of small droplets on rough substrates, where the Darcy precursor film plays a role similar to the precursor film on a smooth substrate. Their findings are consistent with the more general theoretical treatment of \citet{Starov2002,Starov2002b, Starov2003} of small droplet spreading on rough substrates, and subsequent experimental investigations of the same \citep{Grzelakowski2009,Gambaryan2014,Dorbolo2021}. Notably, spreading on deep, porous substrates ($h_2\gtrsim h,\ell_c$) does not yield the same universal behavior~\citep{Davis1998,Starov2002c,Starov2003,Frank2012,Chebbi2021}. In the same vein, we note that the logarithmic correction \cref{eq:logcorrection} to Hoffman's scaling depends on the microstructure of the droplet-film transition on a smooth surface. To our knowledge, evidence for such a logarithmic correction has yet to be observed for spreading on rough substrates.}

\subsection{Existing Models for Total Wetting by Large Droplets}\label{sec:existingresults_large}
For large droplet \dave{spreading} ($R\gg \ell_c$), \dave{the most widely-accepted} model is that of \citet{Lopez1976}, who apply a lubrication approximation to model the interior of the droplet, away from the apparent contact line. The height profile $h$ thus satisfies
\begin{equation}\label{eq:lubrication}
    \partial_t h(r) = -\frac{1}{3\mu}\nabla\cdot\big(h^3\nabla(\sigma \nabla^2 h - \rho g h)\big).
\end{equation}
Lopez \textit{et al.} proceed by supposing that surface tension is negligible in the relatively flat droplet interior \dave{(see \cref{fig:diagram}(\textit{c}))}. The resulting physics is then exactly that of a viscous gravity current, as discussed in \cref{sec:intro}, giving rise to the scaling $R_1\sim V^{3/8}t^{1/8}$. 

\dave{Subsequent work has adapted this gravity-driven model in various ways. \citet{Ehrhard1991} suggested that gravitational overpressure is resisted by edge-localized energy dissipation, rather than the bulk dissipation of \citet{Lopez1976}, and so recovered a scaling $R_1\sim t^{1/7}$. Subsequent authors have argued that bulk dissipation should be dominant, and thus that the widely-observed $R_1\sim t^{1/8}$ scaling should emerge \citep{Hocking1992,Hocking1994,Bonn2009}. In a different direction, \citet{Voinov1995,Voinov1999} suggested perturbatively correcting this gravity-driven model to better account for the effect of surface tension near the edge of the droplet. Voinov posits} two distinct regimes in the droplet: a gravity-driven interior, like that of Lopez \textit{et al.}, and a quasi-static meniscus region. The two regimes are connected by asymptotically matching height profiles at their interface, \dave{and} Voinov uses the contact angle predicted by the fully quasi-equilibrium droplet profile \cref{eq:drop_profile} as a boundary condition for the outer region.

\dave{It has been observed by \citet{Cazabat1986} that, if a droplet's radius $R_1$ grows to exceed the capillary length during the course of spreading, it transitions rapidly from the small-droplet scaling $R_1\sim t^{1/10}$ to the large-droplet scaling $R_1\sim t^{1/8}$. \citet{Ehrhard1993} observed similar behavior, but argued that their data supports the $R_1\sim t^{1/7}$ large-droplet scaling derived by \citet{Ehrhard1991}. In either case, according to the current understanding of these two regimes, such a transition would entail a rapid change from a quasi-equilibrium, edge-driven flow to a dynamic, overpressure-driven flow. We revisit this problem in \cref{sec:priordata}, where we see that the model presented here rationalizes the results of \citet{Cazabat1986} with an edge-driven spreading mechanism valid for both small and large drops.}

\subsection{Existing Models for Partial Wetting on Smooth Substrates}\label{sec:existingresults_partial}
\dave{While the total wetting regime remains our primary focus, it is worth outlining the close connection between the spreading of totally wetting droplets (with spreading parameter $S>0$) and the relaxation to equilibrium of {partially wetting} droplets (with $S<0$). Droplets in both regimes exhibit a leading order geometry as described in \cref{sec:geometry}, as well as unbalanced contact line forces that drive the droplet spreading. These similarities suggest that the present model can be adapted to describe the relaxation to equilibrium of partially wetting droplets, a case we return to in \cref{sec:priordata}.}

\dave{Partially wetting droplets do not generally give rise to precursor films~\citep{Bonn2009}, so are marked by a well-defined contact line where the liquid, substrate, and ambient vapor all meet. Such droplets tend toward a static equilibrium as they spread, in which the drop meets the substrate at a contact angle $\theta_\mathrm{eq}>0$ prescribed by Young's Law \cref{eq:eq_contactangle}. 
Various models have been proposed to rationalize the spreading of partially wetting drops as they relax into equilibrium~\citep{deGennes1990,BrochartWyart1992,Petrov1992,deRuijter1997,deRuijter1999,Eggers2005}. The most recent and comprehensive work on the subject is that of \citet{Durian2022}, which extends the model of \citet{deRuijter1999} to account for droplets of arbitrary horizontal radius.}

\dave{Consistent with prior literature, Durian \emph{et al.} posit that spreading is driven by either capillary or gravitational forces, depending on the horizontal radius of the drop. As in our own model, however, they posit that spreading is resisted by a combination of \emph{bulk} and \emph{edge-localized} energy dissipation, and thereby recover smooth transitions between previously-observed scaling laws. For small droplets, they find an equation of the form
\begin{equation}\label{eq:partial_small}
    \left(\alpha R_1^9/V^3 + \kappa R_1^6/V^2\right)\dot{R}_1 = 1 - (R_1/R_{1,\mathrm{eq}})^6
\end{equation}
for system-dependent parameters $\alpha,\kappa>0$. For large droplets, they similarly find
\begin{equation}\label{eq:partial_large}
    \left(\alpha R_1^7/V^3 + \kappa R_1^4/V^2\right)\dot{R}_1 = 1 - (R_1/R_{1,\mathrm{eq}})^4.
\end{equation}
A key point in the derivation of Durian \emph{et al.} is that the form of the edge-localized energy dissipation in the droplet is \emph{not} the same as that prescribed by \citet{Hervet1984} for totally wetting droplets. Following \citet{deRuijter1999}, the edge-localized energy dissipation for a partially wetting droplet is posited to take the form
\begin{equation}\label{eq:dissipation_partial}
    D_\mathrm{edge}\sim \kappa R_1 (\dot{R}_1)^2,
\end{equation}
independent of the contact angle $\theta$. By comparison, recall that Hervet \& de Gennes suggest the form $D_\mathrm{edge}\sim\beta R_1 (\dot{R}_1)^2/\theta$ for totally wetting droplets. We note that the form \cref{eq:dissipation_partial} is consistent with resistance generated by corrugation of the contact line.}

\dave{We revisit the question of partial wetting in \cref{sec:priordata}, where we show that, by adapting our model to account for substrate-dependent contact line forces, one can interpolate between the small-droplet prediction \cref{eq:partial_small} and the large-droplet prediction \cref{eq:partial_large} using a single edge-driven mechanism.}

\subsection{Washburn's Law for Precursor Films on Rough Substrates}\label{sec:washburn}
Precursor films on rough substrates---\dave{which we refer to as Darcy precursor films}---can be understood as viscous flows through a shallow, porous medium \citep{Gambaryan2014}. Such flows are governed by Darcy's law~\citep{Darcy1856,Whitaker1986}:
\begin{equation}\label{eq:darcy}
    \ovl{u} = -\frac{k}{\mu\phi}\left(\nabla p - f\right).
\end{equation}
Here, $\ovl{u}$ is the local mean velocity of the fluid, $k$ is the permeability of the medium, $\phi$ is the porosity (i.e., void fraction) of the medium, and $f$ is the volumetric force \dave{within the precursor film}. 

\dave{Volume conservation} requires that $\ovl{u} \propto 1/r$, and matching \dave{$\ovl{u}|_{R_2}=\dot{R}_2$} yields $\ovl{u} = R_2\dot{R}_2/r$. The volumetric force $f$ is prescribed by the surface tension acting on the \dave{inner and outer boundaries of the precursor film}. To calculate it, we model the film as a ring of independent, radial strands of fluid, each subtending an angle $d\theta$. The total tension applied to such a strand is $S R_2\,d\theta$, where $S$ is the roughness-dependent spreading parameter of our fluid over the given substrate. The tension-per-unit-radius is thus $S R_2(R_2-R_1)^{-1}\,d\theta$. If we suppose that \dave{the} tension is uniformly distributed across the cross section of fluid at radius $r$, then the volumetric force is
\begin{equation}\label{eq:filmforce}
    f = \frac{SR_2(R_2-R_1)^{-1}\,d\theta}{h_2r\,d\theta} = \frac{\sigma sR_2/r}{R_2-R_1},
\end{equation}
\dave{where we define} $s = S/h_2\sigma$ for convenience. \dave{Finally, if} one neglects the pressure gradient in \cref{eq:darcy}, one finds the following variant of Washburn's law \citep{Washburn1921}:
\begin{equation}\label{eq:R2_bad}
    \dot{R}_2 = \frac{ks}{\phi}\,\frac{1}{R_2-R_1},\qquad \Delta R = R_2-R_1\sim \sqrt{ks t/\phi},
\end{equation}
after making the approximation $\dot{R}_2\gg \dot{R}_1$ and \dave{adopting our non-dimensionalization}. \dave{In practice, while the particular scaling $\Delta R\sim t^{1/2}$ can be achieved with certain combinations of liquids and substrates, previous authors have observed a range of exponents between $0.25$ and $0.5$ \citep{Cazabat1986,Dorbolo2021}.} We will build upon this model through consideration of \dave{gravitational} overpressure from the droplet bulk, which will play a critical role in rationalizing our experimental results. \dave{In \cref{sec:priordata}, we revisit the problem in order to demonstrate how the $R_2\sim t^{3/10}$ spreading reported by \citet{Dorbolo2021} can be rationalized with the same model.}


\section{Experiments: Large Droplets on Rough Substrates}\label{sec:experiment}
\dave{We investigate the spreading of a liquid droplet and its associated \dave{Darcy} precursor film over a rough surface. We focus here on the total wetting of roughened glass surfaces by large droplets ($R_1\gg\ell_c$) of silicone oil. As we discuss in \cref{sec:model}, this case allows us to most clearly distinguish our proposed spreading model from the existing gravity-driven paradigm for large droplets \citep{Lopez1976}. We will apply our model to rationalize existing datasets for small droplets, smooth substrates, and partial wetting in \cref{sec:priordata}.}

\subsection{Methodology}
\begin{figure}
    \centering
    \includegraphics[width=\linewidth]{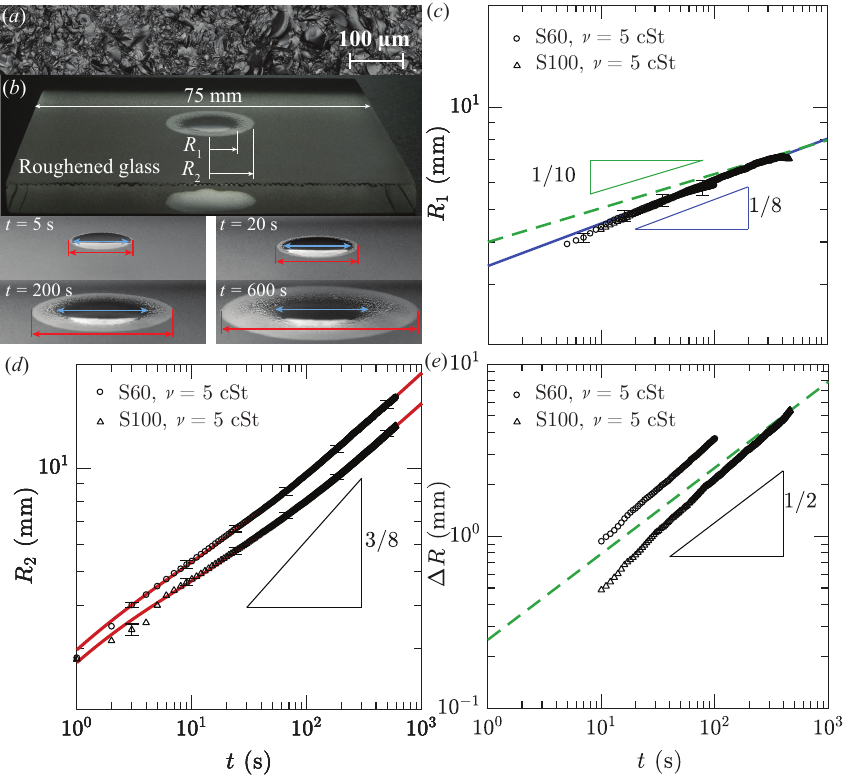}
    \caption{\dave{\textbf{(a)} Confocal microscope scan of the surface S60, a borosilicate glass square sanded with 60 grit silicone carbide lapping compound. \textbf{(b)} Annotated image of a silicone oil droplet wetting the surface S100, sanded instead with 100 grit compound. Shown below are several snapshots of the spreading process, with the drop diameter $2R_1$ (blue arrows) and precursor film diameter $2R_2$ (red arrows) indicated. \textbf{(c--e)} Time evolution of the drop radius $R_1$, the film radius $R_2$, and their difference $\Delta R = R_2-R_1$, for a $5$~\SI{}{\micro\liter} silicone oil droplet ($\nu = 5$~cSt, $\sigma = 20$~mN~m$^{-1}$, $\rho = 0.913$~g~cm$^{-3}$) wetting both S60 and S100. The solid blue line in \textbf{(c)} represents the roughness-independent scaling $R_1 \sim t^{1/8}$ predicted for large droplets ($R_1\gg\ell_c$), and the dashed green line represents the scaling $R_1\sim t^{1/10}$ predicted for small droplets ($R_1\ll\ell_c$). These observations are consistent with our experiments being in the large-droplet regime. The solid red lines in \textbf{(d)} are fits of $R_2$ measurements, equal to $R_2(t) = At^{3/8}/[\log(Bt+C)]^{1/2}$ for surface-dependent parameters $A,B,C$. For the surface S100, the fitting coefficients are 
    $A = 2.99$ mm s$^{-3/8}$, $B = 0.63$ s$^{-1}$, and $C = 2.62$. For the surface S60, the fitting coefficients are $A = 3.80$ mm s$^{-3/8}$, $B = 1.33$ s$^{-1}$, and $C = 3.79$. We rationalize the values of these parameters in \cref{sec:estimates}. We observe that neither curve adheres to the $t^{3/8}$ power law recovered from neglecting log factors in this fit. Finally, the dashed line in \textbf{(d)} represents the scaling $\Delta R(t) \sim t^{1/2}$ predicted by Washburn's law of wicking, which is inconsistent with our observations. We note that previous experiments of viscous droplet spreading also report strong deviations from Washburn's law \citep{Cazabat1986,Dorbolo2021}. We rationalize these deviations in \cref{sec:priordata}}.}
    \label{fig:2}
\end{figure}

Rough surfaces were made by hand-lapping $75\times75$ mm$^2$ square sections of borosilicate glass on a slurry of course-grain silicone-carbide lapping compound and water. Two types of lapping compound with different grain sizes were used to make surfaces of different roughness: the first surface (S60) was sanded with 60 grit (or 250 \SI{}{\micro\meter}) compound, and the second surface (S100) was sanded with 100 grit (120 \SI{}{\micro\meter}). The resulting surfaces were visibly `frosted' (or opaque). The surfaces are characterized by their permeability $k$ and porosity (i.e., void fraction) $\phi$. It is shown that S60 and S100 have $k/\phi = 3.28\times 10^{-5}$~cm$^2$ and $k/\phi = 2.03\times 10^{-5}$~cm$^2$, respectively, implying the pore size decreases as the grit of the lapping compound increases (or the average size of the lapping aggregate decreases). A laser-scanning confocal microscope (VK-X250, Keyence) was used to visualize the topography of the roughened glass substrates. A sample 2-D projection of a microscope scan for S60 is shown in \cref{fig:2}(\textit{a}). Surface asperities exhibit a hierarchy of length scales from tens of nanometers to hundreds of micrometers. The areal footprint of the large-scale asperities observed in \cref{fig:2}(\textit{a}) for S60 are comparable to the derived pore size $k/\phi = 3.28\times 10^{-5}$~cm$^2$. \dave{The root-mean-squared (RMS) roughness height for the S60 surface measured using the confocal miscroscope was 7.6~\SI{}{\micro\meter} and the average lateral length scale (an auto-correlation length) was 56.7~\SI{}{\micro\meter}}. \dave{Due to limitations in equipment availability, we were unable to make the same measurement for S100. Since the lapping process was the same, however, we can scale by the ratio of (square root) permeabilities, $0.79\approx 45$~\SI{}{\micro\meter}~$/$~57~\SI{}{\micro\meter}, to estimate that the RMS roughness height is 6.0~\SI{}{\micro\meter} and the lateral length scale is 44.9~\SI{}{\micro\meter} for S100.}

\subsection{Experimental Results}

\dave{In our primary experiments, a} $V = 5$~\SI{}{\micro\liter} drop of silicone oil (63148-62-9, Millipore Sigma), \dave{with $\sigma = 20$~mN~m$^{-1}$, $\rho = 0.913$~g~cm$^{-3}$, $\mu = 4.6$~mPa~s (or $\nu=\mu/\rho = 5.0$ cSt)}, and initial radius $R_1(t=0)\approx 3.0$~mm, was gently placed on the roughened glass substrate. Here, the capillary length $\ell_c$ is 1.5~mm and the capillary velocity \dave{is $u_c=\sigma/\mu = 4.3$~m~s$^{-1}$. We note that the silicone oils we use are non-volatile at normal (ambient) temperature and pressure, with `no detectable vapor pressure' \citep{ecetoc2011}, and they totally wet borosilicate glass \citep{Dorbolo2021}. A secondary series of experiments with liquids of higher viscosity is reported in \cref{sec:similarity}, and supports the conclusions drawn from our primary experiments.} 

The shape of the drop as it wet the substrate was monitored using a 4~Mpixel digital camera (Lumix GH5, Panasonic) at an image acquisition rate of 1~Hz, with a diffuse spot light for backlit illumination. The camera was positioned at an acute angle above the horizontal plane to promote detection of the droplet bulk from reflection. A sample image of the droplet on S60 is shown in \cref{fig:2}(\textit{b}) with the droplet radius $R_1$ and precursor film radius $R_2$ labelled. An edge detection algorithm developed in MATLAB (MathWorks) was used to extract the droplet bulk radius $R_1$ and precursor film radius $R_2$ \dave{from each image}. \dave{Namely, the image background is first subtracted to enhance contrast with the spreading liquid. Images are then binarized using the method of \citet{Otsu1979}, which adequately discriminates the bright precursor film from the relatively dark background, yielding an estimate of $R_2$. A similar procedure is used to extract the droplet radius $R_1$, after inverting the image to isolate the darker droplet bulk. Measurements of the droplet diameter $2R_1$ and precursor film diameter $2R_2$ are annotated in blue and red, respectively, on the sample snapshots in \cref{fig:2}(\textit{b}).} 

\dave{At least five} experimental trials, each consisting of 600 images taken over 600~s, were \dave{performed for both} roughened glass surfaces. The average values of $R_1(t)$ and $R_2(t)$ across the repeat trials are shown in Figs.~\ref{fig:2}(\textit{c},~\textit{d}). Error bars represent the standard deviation in the measurements of $R_1(t)$ and $R_2(t)$ across the repeated trials. These repeatability errors are no larger than $\pm 5$\% for all instances of time $t$. Sample images of the spreading droplet on S60 at different times $t$ are shown above Figs.~\ref{fig:2}(\textit{c},~\textit{d}). This imaging setup is ideal for capturing droplet spreading on our opaque, roughened surfaces, where there is clear contrast between the droplet, film, and substrate, as illustrated in \cref{fig:2}. However, it is less effective for visualizing droplet spreading on smooth, transparent glass substrates, where contrast is significantly reduced. Measurements of $R_2$ for spreading on a smooth substrate require techniques with much larger spatial resolution, given that $h_2 \sim \mathcal{O}(10$ \r{A}) in this case. We do not pursue such measurements here, as there are several previous investigations of droplet spreading on smooth substrates using methods such as ellipsometry or x-ray reflectivity that can accurately resolve both radii \citep{Bascom1964,Daillant1988,Cazabat1997,Popescu2012}. \dave{In \cref{sec:priordata}, we will revisit existing datasets of \citet{Cazabat1986} and \citet{deRuijter1999} for droplet spreading over smooth substrates.}

\cref{fig:2}(\textit{c}) demonstrates the evolution of $R_1$ with time, and \cref{fig:2}(\textit{d}) shows \dave{the evolution} of $R_2$. Measurements for the surfaces of different roughness are represented with different symbols (black triangles for S100, black circles for S60). For all observed times $t$, the droplet radius $R_1$ is large ($R_1 \gg \ell_c = 1.5$~mm). Consistent with previous studies \citep{Lopez1976,Cazabat1986,Voinov1999}, fits of $R_1(t)$ in this large-droplet regime demonstrate a power-law trend with an exponent of $1/8$, independent of the surface roughness. \dave{This fit is shown with a solid blue curve in \cref{fig:2}(\textit{c}). A dashed green curve with a power-law exponent of $1/10$, the predicted scaling for small droplets~\citep{Voinov1976,Tanner1979,Hervet1984}, is shown for comparison}. 


For all values of $t$ and $k$, $R_2$ exhibits excellent agreement with a function of the form $R_2(t) = At^{3/8}/[\log(Bt+C)]^{1/2}$. These fits are shown with red solid curves in \cref{fig:2}(\textit{d}) for both surfaces. For the surface S100, \dave{the parameters} 
$A = 2.99\text{ mm}\text{ s}^{-3/8}$, $B = 0.63 \text{ s}^{-1}$ and $C = 2.62$ produce good overlap with the measurements; the relative root-mean-square deviation between the measurements and the fit is 0.68\%. For the surface S60, the fitting coefficients $A = 3.80 \text{ mm}\text{ s}^{-3/8}$, $B = 1.33 \text{ s}^{-1}$, and $C = 3.79$ yield a root-mean-square deviation of 0.39\%. We note that the numerical value of $C$ is negligible beyond $t\gtrsim 10$~s; we rationalize the observed scalings of $R_1$ and $R_2$ in \cref{sec:model} and the observed values of $A$ and $B$ in \cref{sec:estimates}. \dave{For the sake of comparison, we show in \cref{fig:2}(\textit{e}) that neither set of experiments satisfies the $\Delta R = R_2-R_1\sim t^{1/2}$ scaling predicted by the classic version of Washburn's law (see \cref{sec:washburn}).}

\dave{A few features of \cref{fig:2}(\textit{c},~\textit{d}) merit further comment. First, we note that the droplet and film radii can only be clearly distinguished beyond $t\sim 5$--$10$~s, indicating that the Darcy precursor film is not fully developed until that time. Similarly, the droplet depth becomes comparable to the roughness scale when $t\gtrsim 400$~s for S100 and when $t\gtrsim 100$~s for S60. We believe these limitations account for the slight deviations of both $R_1(t)$ and $R_2(t)$ from predictions for $t\lesssim 5$~s and those of $R_1(t)$ for $t\gtrsim 400$~s. Next,} the evolution of $R_2$ depends on the substrate roughness $k$; as the substrate roughness decreases from S60 to S100, $R_2$ also decreases for all $t$. Physically, this \dave{trend} reflects the fact that resistance to flow decreases with greater substrate permeability. \dave{Finally}, \cref{fig:2}(\textit{d}) highlights the need for the $(\log t)^{1/2}$ \dave{term} in the denominator of the scaling for $R_2(t)$; \dave{the spreading is not well described by} a $t^{3/8}$ scaling. 


\subsection{Dynamic Similarity}\label{sec:similarity}

\dave{We here present the results of a secondary series of experiments, in which we compare the spreading rates of silicone oils of different viscosities on the same rough substrate S100. We perform experiments using $\nu = 10$~cSt and $\nu=50$~cSt silicone oil, to complement our data for $\nu = 5$~cSt silicone oil. The material properties of all three fluids are listed in \cref{tab:mat-prop}.  The same experimental setup is used, but with longer acquisition times. For the $\nu = 10$~cSt silicone oil, 600 images are collected at an acquisition rate of 0.5 frames per second, equivalent to 20 minutes of data collection. For the $\nu = 50$~cSt silicone oil, 600 images are collected at an acquisition rate of 0.1 frames per second, equivalent to 100 minutes of data collection.}

\begin{table}
    \begin{center}
        \def~{\hphantom{0}}
        {\begin{tabular}{cccccc}
        
            $\nu$ (cSt) & $\rho$ (kg m$^{-3}$) & $\mu$ (cP) & $\sigma$ (mN m$^{-1}$) &$\ell_c$ (mm) & $u_c$ (m s$^{-1}$)\\

            5  & 913 & 4.6  & 20 & 1.49 & 4.38\\
            10 & 930 & 9.3  & 20 & 1.48 & 2.15\\
            50 & 960 & 48.0 & 20 & 1.46 & 0.42\\
            
        \end{tabular}}
        \caption{\dave{Material properties of the different test fluids in \cref{sec:similarity}.}}
        \label{tab:mat-prop}
    \end{center}
\end{table}

\begin{figure}
    \centering
    \includegraphics[width=\linewidth]{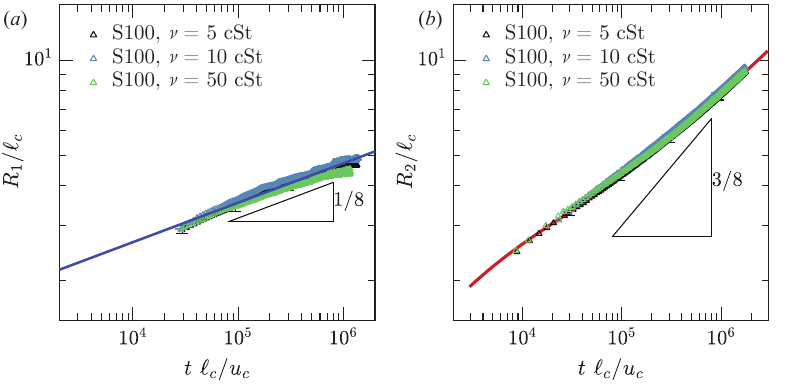}
    \caption{\dave{Evolution of \textbf{(a)} the droplet radius $R_1(t)$ and \textbf{(b)} the Darcy precursor film radius $R_2(t)$ for silicone oils of kinematic viscosity $\nu = 5$~cSt, $10$~cSt, and $50$~cSt on the surface S100. When non-dimensionalized with respect to the capillary length $\ell_c$ and plotted against the dimensionless time $\hat{t} = t\ell_c/u_c$, all three measurements of $R_1$ align closely with a single $R_1(\hat{t})/\ell_c \sim \hat{t}^{1/8}$ curve (blue), and all three measurements of $R_2$ align closely with a single $R_2( \hat{t}) = A \hat{t}^{3/8}/[\log(B \hat{t}+C)]^{1/2}$ curve (red), where $A = 0.10$, $B = 2.0 \times 10^{-4}$, and $C = 2.50$.}}
    \label{fig:3}
\end{figure}

\dave{\cref{fig:3}(\textit{a},~\textit{b}) illustrate the evolution of the droplet radius $R_1(t)$ and the Darcy precursor film radius $R_2(t)$, respectively, for silicone oils of viscosity $\nu = 5$~cSt, $10$~cSt, and $50$~cSt on the roughened surface S100. When the radii $R_1(t)$ and $R_2(t)$ are normalized with respect to the capillary length $\ell_c$ and plotted with respect to a dimensionless time $\hat{t} =t \ell_c/u_c$, the measurements overlap closely, indicating that the spreading dynamics are dynamically similar. As before, the droplet radius exhibits a $R_1(\hat{t})/\ell_c\sim \hat{t}^{1/8}$ scaling, and the Darcy precursor film fits follows the trend ${R}_2( \hat{t})/\ell_c = A \hat{t}^{3/8}/[\log(B \hat{t}+C)]^{1/2}$. Fits of the data yield dimensionless fitting coefficients of $A = 0.10$, $B = 2.0 \times 10^{-4}$ and $C = 2.50$. Because all three experiments are dynamically similar to the $\nu = 5$~cSt case, the discussion in \cref{sec:estimates} can be used to rationalize these coefficients.}

\section{Theoretical Model}\label{sec:model}
\dave{We proceed by developing a theoretical model of viscous droplet spreading that both rationalizes
our own experimental observations and provides insight into prior experimental studies}. The core feature of our model is that the droplet maintains a quasi-equilibrium state as it evolves, with curvature and hydrostatic pressure in balance according to the calculated profile \cref{eq:drop_profile}. A similar assumption was applied by \citet{Tanner1979}, \citet{Hervet1984}, and others in the small droplet limit, where the droplet maintains the shape of a spherical cap, \dave{thereby minimizing surface area}. We remark that the height profile we deduce from this balance is not novel \citep{Tanner1979,Voinov1999}; the novelty of our approach is the use of \cref{eq:drop_profile} as an evolving quasi-equilibrium, driven by unbalanced interfacial forces acting along the droplet's edge. We discuss in turn the dynamics of the spreading droplet and the \dave{Darcy precursor film that precedes it when it spreads on a rough substrate}.

\subsection{Dynamics of the Droplet Edge}\label{sec:contactline}
We \dave{first derive} an expression for the speed at which the droplet's \dave{edge} advances, using a simple scaling argument. The present section simplifies and generalizes the argument of \citet{Hervet1984} presented in \cref{sec:existingresults_small}.

\dave{We suppose here that viscous dissipation is not necessarily confined to the vicinity of the droplet edge, as posited by \citet{Hervet1984}, but that some component of the dissipation may occur throughout the droplet bulk. The total dissipation may thus be expressed as
\[D = D_\mathrm{edge} + D_\mathrm{bulk}.\]
Lifting the assumption of edge-localized dissipation is natural in the Stokes-flow limit, where one expects instantaneous equilibration of the flow field throughout the droplet bulk; indeed, splitting the total energy dissipation into `edge' and `bulk' contributions was previously suggested by \citet{deRuijter1999} and \citet{Durian2022} for the case of partial wetting, and bulk dissipation was assumed in the gravity-driven model of \citet{Lopez1976}. For a simple, concrete flow field that exhibits the properties we seek, one could imagine that, away from the edges, a large droplet would evolve as a slow Poiseuille-type current of the form
\[u(r,z) \sim U\frac{r}{R_1}\frac{z}{h_1}\left(2-\frac{z}{h_1}\right).\]
We emphasize that we do \emph{not} prescribe such a flow profile. The model we present relies only on scaling arguments, and thus rationalizes the observed spreading rates of viscous droplets without detailed knowledge of the droplets' flow profiles.}

\dave{Suppose $\Phi$ is the average local rate of energy dissipation throughout the droplet bulk, and $U\sim\dot{R}_1$ is the rate of spreading. Then the bulk dissipation rate is given by}
\dave{\begin{equation}\label{eq:dissipation}
D_\mathrm{bulk} \sim \alpha V\Phi \sim \alpha \mu h_1 R_1^2 \frac{U^2}{h_1^2} \sim \frac{\alpha\mu R_1^2I_1(R_1)U^2}{(I_0(R_1)-1)\tan\theta},
\end{equation}
for some constant $\alpha>0$. The above formula} is deduced by substituting\footnote{\dave{A similar result would follow if we instead used the full expression \cref{eq:volume} to estimate the volume in \cref{eq:dissipation}; in particular, the asymptotic limits of our model would remain unchanged.}} $V\sim h_1R_1^2$ and $h_1=h|_{r=0}$. We remark that $V\sim h_1R_1^2$ holds for both small and large droplets.

\dave{An expression for $D_\mathrm{edge}$ can be derived using the analytical argument of \citet{Hervet1984}, presented in \cref{sec:existingresults_small}. For the sake of completeness, however, we illustrate how the form of $D_\mathrm{edge}$ can be deduced from a similar scaling argument as that of $D_\mathrm{bulk}$. If dissipation occurs within a distance $L$ of the edge, then it occurs over the solid of rotation carved out by a triangle of height $H=L\tan\theta$ and area $A = \frac{1}{2}L^2\tan\theta$; this solid has a volume $V_\mathrm{edge}\sim R_1HL$, and we expect the total dissipation within it to take the form
\begin{equation}\label{eq:dissipation_edge}
    D_\mathrm{edge}\sim\beta V_\mathrm{edge}\Phi_\mathrm{edge} \sim \beta \mu R_1HL\frac{U^2}{H^2} = \beta \mu R_1\frac{U^2}{\tan\theta},
\end{equation}
for some constant $\beta>0$. Setting $\beta = \log(L/a)$ recovers the result of Hervet and de Gennes (see \cref{sec:existingresults_small}). We remark that, \emph{a priori}, there might also be a contribution of the form~\cref{eq:dissipation_partial} proposed by \citet{deRuijter1999} and \citet{Durian2022}, dependent only on the spreading speed $U$ and the arclength $2\pi R_1$ of the apparent contact line and \emph{independent} of the apparent contact angle $\theta$. Such a contribution might correspond to micro-physics at the droplet-film interface, at length scales $\ell\ll L$; for the analogous case of partially wetting drops on rough substrates, for instance, one expects resistance to arise from microscopic corrugation of the contact line~\citep{deGennes2003}. In all, the edge-localized dissipation would take the more general form
\[D_\mathrm{edge}\sim \beta\mu R_1\frac{U^2}{\tan\theta} + \kappa\mu R_1 U^2,\]
for some constant $\kappa>0$. In any case, the total energy dissipation $D$ takes the form
\begin{equation}\label{eq:totaldissipation}
    D = D_\mathrm{bulk} + D_\mathrm{edge} \sim \left(\alpha \frac{R_1^2I_1(R_1)}{(I_0(R_1)-1)\tan\theta} + \beta \frac{R_1}{\tan\theta} + \kappa R_1\right)\mu U^2.
\end{equation}
One can simplify this expression in our limits of interest. First, we assume that $\beta R_1/\tan\theta \gg \kappa R_1$, because $\theta\ll 1$. This assumption can otherwise be justified by considering that, in the small droplet limit, the $\beta$ term must be dominant in order to recover the widely-corroborated scaling predictions of \citet{Hoffman1975}, \citet{Voinov1976}, and \citet{Tanner1979}. Secondly, the remaining contribution of $D_\mathrm{edge}$ does not change the asymptotic form of $D$ in either the large or small droplet limit. In the limit $R_1\gg \ell_c$, the $\alpha$ term is quadratic in $R_1$ and thus dominates the edge dissipation. In the limit $R_1\ll \ell_c$, both the $\alpha$ and the $\beta$ terms are linear in $R_1$, and one can simply augment $\alpha\mapsto \alpha+\beta$ to account for both. Since we are ultimately interested in these asymptotic limits, we proceed as though edge dissipation is accounted for in the value of $\alpha>0$ for small droplets.}

The work done in moving the contact line is $W \sim \sigma(1-\cos\theta) R_1U$. The droplet's internal energy balance is given by $D\sim W$, which upon rearrangement yields
\begin{equation}\label{eq:degennes}
    \frac{R_1I_1(R_1)}{I_0(R_1)-1}\,\dot{R}_1 = 2\alpha^{-1}(\tan\theta - \sin\theta)\sim \alpha^{-1}\theta^3,
\end{equation}
for $\theta\lesssim\pi/4$ and \dave{the same constant $\alpha>0$ as above}. We employ the asymptotic formulas \cref{eq:asymptotic} to deduce the small- and large-drop limits of this result. For small droplets ($R_1\ll \ell_c$), we have \dave{$R_1I_1(R_1)/[I_0(R_1)-1]\to 2=\mathrm{const}.$}, so \cref{eq:degennes} recovers the $\dot{R}_1\sim \theta^3$ \dave{scaling} of \citet{Hoffman1975}. For large droplets, $R_1I_1(R_1)/[I_0(R_1)-1]\to R_1$, so \cref{eq:degennes} yields the modified scaling $\dot{R}_1\sim \theta^3/R_1$.

Using the expression \cref{eq:volume} for the droplet volume to eliminate $\theta$, we find
\begin{equation}\label{eq:hoffman_u}
\dot{R}_1 =\alpha^{-1}\,\frac{I_0(R_1)-1}{R_1I_1(R_1)}\,\Bigg(\frac{V/\pi}{R_1^2\frac{I_0(R_1)}{I_1(R_1)} - 2R_1}\Bigg)^{3}.
\end{equation}
\dave{We note that the liquid volume that drains into the precursor film (over either smooth or rough substrates) is negligible over the time scales of interest. Indeed, because the roughness (or microscopic) scale $h_2$ is small relative to the droplet depth $h_1$, the liquid volume contained within the film is negligible with respect to that within the droplet:
\begin{equation}\label{eq:Vfilm}
    V_\mathrm{film} = h_2R_2^2\ll h_1R_1^2\sim V,
\end{equation}
noting that the scaling $V\sim h_1R_1^2$ holds for both small and large droplets. Provided \cref{eq:Vfilm} holds, the change in droplet volume $V$ is negligible, so} \cref{eq:hoffman_u} yields an expression for the evolution of the droplet radius\footnote{\dave{We note that the system could alternatively be closed by an equation of the form
\[\dot{V} = -\dot{V}_\mathrm{film}=-2\pi\phi h_1(R_1\dot{R}_1 - R_2\dot{R}_2),\]
where $\phi$ is the porosity of the rough substrate (and $\phi=1$ for smooth substrates). We do not pursue this direction further at present, and instead take $V$ to be constant.}}. 

The physical picture \dave{we propose} can be summarized as follows, for both large and small drops. At all times, hydrostatic pressure and curvature pressure are in balance \dave{throughout the droplet, except in the immediate vicinity of the apparent contact line. At the apparent contact line, interfacial forces act to stretch the droplet uniformly outward.} The work done \dave{by these interfacial forces} is dissipated \dave{inside the droplet}, and \dave{the associated} energy balance determines the rate of \dave{spreading}. 

The basic physical picture is common to both small and large droplets; different scalings arise in the two limits only because the relationship between $V$, $\theta$, and $R_1$ changes according to \cref{eq:volume}. In the small droplet limit $R_1\ll\ell_c$, the asymptotic formulas \cref{eq:asymptotic} show that the right-hand side of \cref{eq:hoffman_u} scales as $V^3/R_1^{9}$, so the evolution equation reduces to the $R_1(t) \sim V^{3/10}t^{1/10}$ scaling of \dave{\citet{Hoffman1975}, \citet{Voinov1976}, and \citet{Tanner1979}}. For large droplets, the right-hand side of \cref{eq:hoffman_u} scales as $V^3/R_1^{7}$, which yields a scaling $R_1(t)\sim V^{3/8}t^{1/8}$ familiar from viscous gravity currents. We summarize the small and large droplet limits of \dave{our} model in \cref{tab:kd}, in both 1-D and 2-D geometries. \dave{Since the derivation presented here depends on scaling arguments, one should expect \cref{eq:hoffman_u} to hold in both limits, but \emph{not} necessarily to yield exact predictions for the small-to-large-droplet transition. We return to the latter question in \cref{sec:priordata}.}

We note that our predicted $R_1\sim V^{3/8}t^{1/8}$ scaling for large droplets carries the same prefactor as that of gravity currents. Indeed, reintroducing units to \dave{our scaling yields}
\[R_1 \sim \ell_c(V/\ell_c^3)^{3/8}(t u_c/\ell_c)^{1/8} = \ell_c^{-1/4}u_c^{1/8}V^{3/8}t^{1/8} = (\rho g/\mu)^{1/8}V^{3/8}t^{1/8}.\]
\dave{Notably, despite the dynamics of our model being driven by interfacial forces,  $\sigma$ necessarily drops out from the final scaling. We can rationalize this phenomenon as follows}. Recall that the horizontal contact line force per unit arclength is $f\sim\sigma\theta^2$. \dave{If one fixes the droplet radius $R_1$ and depth $h_1$, then the equilibrium droplet profile \cref{eq:drop_profile} shows that the contact angle scales with $\sigma$ as $\theta\sim h_1/\ell_c\sim h_1(\rho g/\sigma)^{1/2}$. Thus, the total interfacial} force scales as
\[f\sim \sigma\theta^2\sim \rho g h_1,\]
independent of $\sigma$. 
\dave{As a consequence, one cannot distinguish our edge-driven spreading model from traditional gravity currents on the basis of $R_1(t)$ alone}. We distinguish between the two models in the following \dave{subsection}, by testing their respective predictions for the evolution of \dave{a Darcy precursor film when a droplet spreads on a rough substrate}.


\begin{table}
  \begin{center}
\def~{\hphantom{0}}
  \begin{tabular}{cccc}
      Small Droplet, 2-D  & Large Droplet, 2-D   &   Small Droplet, 1-D & Large Droplet, 1-D \\($R_1\ll\ell_c$)  & ($R_1\gg\ell_c$)  &   ($R_1\ll\ell_c$) & ($R_1\gg\ell_c$) \\\hline
       $\dot{R}_1\sim \theta^3$  & $\dot{R}_1\sim  R_1^{-1}\theta^3$ & $\dot{R}_1\sim\theta^3$ & $\dot{R}_1\sim R_1^{-1}\theta^3$\\[3pt]
       $R_1\sim V^{3/10}t^{1/10}$   & $R_1\sim V^{3/8}t^{1/8}$  & $R_1\sim V^{3/7}t^{1/7}$ & $R_1\sim V^{3/5}t^{1/5}$\\[3pt]
       $\theta\sim V^{1/10}t^{-3/10}$   & $\theta\sim V^{1/4}t^{-1/4}$  & $\theta\sim V^{1/7}t^{-2/7}$ & $\theta\sim V^{2/5}t^{-1/5}$\\[3pt]
       $h_1\sim V^{2/5}t^{-1/5}$   & $h_1\sim V^{1/4}t^{-1/4}$  & $h_1\sim V^{4/7}t^{-1/7}$ & $h_1\sim V^{2/5}t^{-1/5}$\\[10pt]\multicolumn{4}{c}{Wicking-Driven Darcy Precursor Films}\\[-4pt]\hline
       $\Delta R\sim k^{1/2}s^{1/2}t^{1/2}$  & $\Delta R\sim k^{1/2}s^{1/2}t^{1/2}$ & $\Delta R\sim k^{1/2}s^{1/2}t^{1/2}$ & $\Delta R\sim k^{1/2}s^{1/2}t^{1/2}$\\[10pt]\multicolumn{4}{c}{Overpressure-Driven Darcy Precursor Films}\\[-4pt]\hline
        $R_2 \sim \frac{k^{1/2}V^{-1/10}t^{3/10}}{\sqrt{\log kV^{-1/5}t^{3/5}}}$ & $R_2 \sim \frac{k^{1/2}V^{1/8}t^{3/8}}{\sqrt{\log kV^{1/4}t^{3/4}}}$ & $\Delta R \sim k^{1/2}V^{-1/7}t^{2/7}$ & $\Delta R \sim k^{1/2}V^{1/5}t^{2/5}$\\
  \end{tabular}
  \caption{Scalings predicted by our model in various limits of interest, for both 1-D and axisymmetric 2-D systems. \dave{Here, $R_1$, $h_1$, and $\theta$ are the horizontal radius, depth, and apparent contact angle of the droplet bulk, $R_2$ is the radius of the Darcy precursor film on a rough substrate, and $\Delta R=R_2-R_1$. The scalings reported for $\dot{R}_1$, $R_1$, $h_1$, and $\theta$ are predicted to hold for droplets totally wetting either smooth or rough substrates, but we clarify that we do not predict the spreading rate of a microscopic precursor film on a smooth substrate.} The $\dot{R}_1\sim\theta^3$ scaling of \citet{Hoffman1975} and \citet{Hervet1984} arises in the small droplet limit in both 1-D and 2-D systems. The $R_1\sim V^{3/10}t^{1/10}$ and $\theta\sim V^{1/10}t^{-3/10}$ laws of \citet{Voinov1976} and \citet{Tanner1979} arise in the small droplet limit in 2-D. The $R_1\sim V^{3/7}t^{1/7}$ scaling of \dave{\citet{Tanner1979} and} \citet{Mchale1995} arises in the small droplet limit in 1-D. The $\Delta R\sim t^{1/2}$ law of \citet{Washburn1921} arises in the limit of tension-driven film dynamics. The $R_1\sim V^{3/8}t^{1/8}$ scaling observed by \citet{Lopez1976} arises for large droplets in 2-D.}
  \label{tab:kd}
  \end{center}
\end{table}

\subsection{Dynamics of a \dave{Darcy} Precursor Film on a Rough Substrate}\label{sec:R2-scaling}
We model the \dave{Darcy} precursor film as a viscous, two-dimensional flow through a homogeneous, porous medium, subject to Darcy's law \cref{eq:darcy}. Here, we modify the statement of Washburn's law \cref{eq:R2_bad} to account for the pressure gradient across the precursor film, associated with overpressure in the droplet bulk. \dave{Similar calculations were carried out by \citet{Clarke2002} and \citet{Starov2002,Starov2003} for small droplets on deep and shallow porous substrates, respectively, but we here give a more detailed account of the time-evolving overpressure within the droplet}.

Following the \dave{same argument as} \cref{sec:washburn}, we know that \dave{the local mean velocity $\ovl{u}$ and volumetric force $f$ within the precursor film must satisfy $\ovl{u},f\propto 1/r$. These are two of the three terms appearing in Darcy's law \cref{eq:darcy}, so we deduce that the third term must have the same dependence, i.e., $\nabla p\propto 1/r$}. Matching the pressure conditions at the leading and trailing edge of the film, $p|_{r=R_2} = 0$ and $p|_{r=R_1}=-\sigma\nabla^2h(R_1)$, we find
\[p = -\sigma(\nabla^2 h)(R_1)\,\frac{\log(r/R_2)}{\log(R_1/R_2)} =\frac{\sigma V/\pi}{R_1^2 -2R_1\frac{I_1(R_1)}{I_0(R_1)}}\,\frac{\log(r/R_2)}{\log(R_1/R_2)}.\]
Inserting this expression into \cref{eq:darcy} \dave{and using the expression \cref{eq:filmforce} for $f$, we deduce a modified evolution equation} for the radius of the precursor film:
\begin{equation}\label{eq:R2}
    R_2\dot{R}_2 = \frac{k}{\phi}\left(\frac{V/\pi}{(R_1^2-2R_1\frac{I_1(R_1)}{I_0(R_1)})\log(R_2/R_1)} + \frac{sR_2}{R_2-R_1}\right).
\end{equation}
Depending on surface chemistry, the system obeys one of two different dynamics. First, in the limit where surface tension dominates (roughly, if $\sigma s\gg \rho g\ell_c$), we recover the \dave{classic} variant of Washburn's law, derived in \cref{sec:washburn}. In the limit where hydrostatic pressure dominates ($\sigma s\ll \rho g\ell_c$), we approximate $\log (R_2/R_1)\sim \log R_2$ in order to write
\begin{equation}\label{eq:logthing}
    \log(R_2) R_2\dot{R}_2\sim \frac{k}{\phi}\frac{V}{R^2_1 - 2R_1\frac{I_1(R_1)}{I_0(R_1)}}.
\end{equation}
In the large droplet limit $R_1\gg\ell_c$, for instance, we find $\log(R_2) R_2^2 \sim (k/\phi)\alpha^{1/4}V^{1/4} t^{3/4}$. In general, if $y^n\log y = x$, then $y = \exp(n^{-1}W(nx))$, where $W$ is the Lambert $W$-function \citep{Corless1996}. Since $W(x)\sim \log x - \log\log x$ for large arguments, we find
\begin{equation}\label{eq:logthing_afterapprox}
    R_2(t) \sim \frac{(k/\phi)^{1/2}\alpha^{1/8}V^{1/8}t^{3/8}}{\sqrt{\log{(k/\phi)\alpha^{1/4}V^{1/4}t^{3/4}}}}\sim \frac{t^{3/8}}{\sqrt{\log{t}}}.
\end{equation}
Our experimental results in \cref{sec:experiment} \dave{are consistent with} this scaling, and \dave{would} appear to disqualify \dave{simpler} power-law scalings. The small droplet limit follows similarly, giving $R_2\sim t^{3/10}/(\log{t})^{1/2}$. All scalings are reported in \cref{tab:kd}.

\dave{We note that the logarithmic correction to our predicted $R_2$ scaling is necessary to rationalize our experimental results in \cref{sec:experiment}. This correction arises from the circular geometry of the system, as it does in the scalings found for coalescence of circular droplets in the work of \citet{eggers_coalescence_1999}. Indeed, the 1-D equivalent of our system---corresponding to the `fluid stripe' experiments of \citet{Mchale1995}---gives a modified power law with no logarithmic factors. }

\dave{Finally, we} remark that the $R_2\sim t^{3/8}/(\log t)^{1/2}$ scaling regime is \emph{not} consistent with the gravity-driven hypothesis introduced by \citet{Lopez1976} and currently understood to underlie the $R_1\sim V^{3/8}t^{1/8}$ scaling of large droplets \citep{Bonn2009,Popescu2012}. Gravity-driven spreading implies a horizontal pressure gradient across the droplet radius, and thus a \dave{relatively low} pressure $p|_{r=R_1}\ll \rho gh_1$ at the edge. By contrast, the $R_2\sim t^{3/8}/(\log t)^{1/2}$ scaling regime arises owing to hydrostatic overpressure $p|_{r=R_1}\approx \rho gh_1$ at the droplet edge, consistent with our quasi-equilibrium model. In demonstrating the existence of an $R_2\sim t^{3/8}/(\log t)^{1/2}$ film spreading regime, our experiments in \cref{sec:experiment} \dave{would thus appear to} distinguish the large-droplet limit of \dave{our edge-driven model} from gravity currents.

\section{Broader Applications: Small Droplets, Smooth Substrates, and Partial Wetting}\label{sec:priordata}
\dave{We proceed by presenting three datasets from the literature, for which the present work may offer a valuable new perspective. In turn, these datasets characterize the spreading of a Darcy precursor film ahead of a small droplet on a rough substrate \citep{Dorbolo2021}, the early-time spreading of partially wetting droplets on smooth substrates \citep{deRuijter1999,Durian2022}, and the transition from small to large droplet regimes on smooth substrates \citep{Cazabat1986}. All droplets under consideration are shallow ($h\lesssim \ell_c$), and the descriptors `small' and `large' indicate how their horizontal radii compare to the capillary length.} 

\begin{figure}
    \centering
    \includegraphics[width=\linewidth]{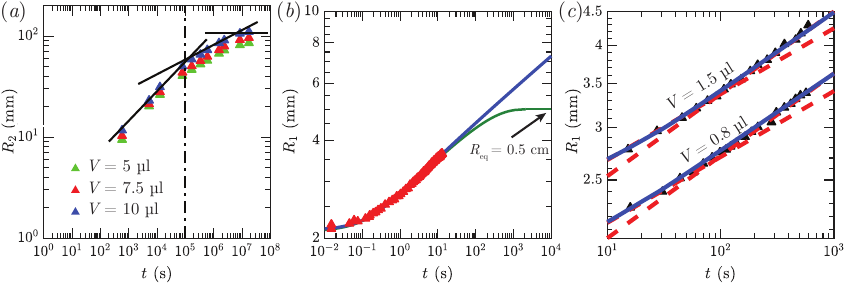}
    \caption{\dave{\textbf{(a)} Downsampled version of the data recorded by \citet{Dorbolo2021} for the spreading of a Darcy precursor film ahead of a small drop of 20~cSt silicone oil on a frosted glass substrate. The radius $R_2$ of the film undergoes three distinct spreading regimes: $R_2\sim t^{3/10}$ for the intermediate time regime ($t\lesssim 10^5$~s) of present interest, $R_2\sim t^{3/20}$ after the droplet bulk sinks into the surface roughness ($10^5$~s~$\lesssim t\lesssim 10^7$~s); and $R_2\sim\mathrm{const}.$ at very late times ($t\gtrsim 10^7$~s). Although Dorbolo originally inferred that the first spreading regime was driven by wicking, with deviations from Washburn's law (\cref{sec:washburn}) caused by irregular roughness elements, we note that the observed scaling $R_2\sim t^{3/10}$ is consistent with our predictions for overpressure-driven Darcy precursor films ahead of small drops. \textbf{(b)} The data recorded by \citet{deRuijter1999} for the spreading of a partially wetting, viscous drop of DBP on a smooth PET substrate, along with two fitted curves reproduced from \citet{Durian2022}: the predicted behavior of the droplet if the equilibrium radius were $R_\mathrm{eq}=0.5$~cm (green); and the predicted early-time spreading behavior, independent of $R_\mathrm{eq}$ (blue). A natural extension of our work to partially wetting drops yields the evolution equation \cref{eq:partialwetting_full}, which reproduces the predictions of \citet{Durian2022} for both large and small drops with a purely edge-driven model. \textbf{(c)} Data reported by \citet{Cazabat1986} for the spreading of viscous drops of PMS on smooth glass, depicting the small-to-large-droplet transition as $R_1$ exceeds the capillary length. Also shown are the asymptotic $R_1\sim t^{1/10}$ and $R_1\sim t^{1/8}$ curves (dashed red) for both $V=1.5$~\SI{}{\micro\liter} and $V=0.8$~\SI{}{\micro\liter}, reproduced from \citet{Cazabat1986}, and the predictions of our own model for both cases (blue). Specifically, we report the predictions of the evolution equation \cref{eq:hoffman_u} augmented with the three dissipation terms present in \cref{eq:totaldissipation}, with the fitted coefficients $\alpha=1$, $\beta=15$, $\kappa = 150$.}}
    \label{fig:priorwork}
\end{figure}

\dave{In \cref{fig:priorwork}(\textit{a}), we reproduce a downsampled version of the data recorded by \citet{Dorbolo2021} for the spreading of a Darcy precursor film ahead of a small drop of silicone oil ($\nu=\mu/\rho = 20$ cSt) on a frosted glass substrate. The glass is sold commercially as `SATINOVO\textsuperscript{\textregistered} MATT'  \citep{SaintGobain2024}, and Dorbolo reported it as having a root-mean-square height variation of $2.34$~\SI{}{\micro\meter} and a characteristic distance between peaks of $34\pm 10$~\SI{}{\micro\meter}, somewhat finer than either of the substrates used in our experiments. The precursor film undergoes three characteristic spreading regimes over $120$ days ($\sim 10^7$~s) of observation. First, it undergoes an intermediate time regime ($t\lesssim 10^5$~s) with a clear distinction between droplet bulk and Darcy precursor film, the latter advancing according to $R_2\sim t^{3/10}$. Dorbolo inferred that this spreading was driven by wicking, with a deviation from Washburn's law (\cref{sec:washburn}) caused by irregular roughness elements. After the droplet bulk sinks into the surface roughness ($10^5$~s~$\lesssim t\lesssim 10^7$~s), the film spreads according to $R_2\sim t^{3/20}$, a scaling previously observed by \citet{Cazabat1986}. Finally, for $t\gtrsim 10^7$~s, the film appears to stops spreading. While the latter two regimes are beyond the scope of our work, the initial spreading regime is consistent with our model.}

\dave{The initial radius $R_1(t=0)$ of the droplet is not reported, but extrapolating the scaling $R_2\sim t^{3/10}$ backwards indicates that $R_1(t=0)< R_2(t=1$~s$)\lesssim 1.5\ell_c$, so it can be assumed that the droplet is relatively small. The prediction $R_2\sim t^{3/10}$ is thus consistent with our own prediction $R_2\sim t^{3/10}/(\log t)^{1/2}$ in the long-time limit for Darcy precursor films driven by small droplets. We note that, while the logarithmic factor was necessary to explain short-time behaviors (particularly $t\lesssim 10^2$~s) in our own experiment, the logarithm grows very slowly for large arguments. One thus expects it to become negligible over the much longer timescale at which the initial spreading regime is observed in Dorbolo's experiment ($\sim 10^5$~s). More generally, we note that the prior experiments of \citet{Cazabat1986} observed a range of exponents for this process between $0.25$ and $0.5$. Our model suggests that these exponents might arise from a combination of overpressure-driven and wicking-driven spreading.}

\dave{We turn now to droplet spreading on smooth substrates. \cref{fig:priorwork}(\textit{b}) depicts the data recorded by \citet{deRuijter1999} for the spreading of partially wetting droplets. Specifically, they studied the spreading of a small droplet of dibutyl phthalate (DBP; $\mu=19.6$~mP~s, $\sigma = 34.3$~mN~m$^{-1}$) on a smooth polyethylene terephthalate (PET) substrate. Under normal conditions, DBP partially wets PET with a very low equilibrium contact angle, and does \emph{not} give rise to a precursor film. This dataset was revisited in the recent work of \citet{Durian2022}, in order to validate their model for the relaxation of partially wetting droplets into equilibrium. In \cref{fig:priorwork}(\textit{b}), we reproduce fitted curves calculated by \citet{Durian2022}. The green curve corresponds to an example choice of equilibrium drop radius $R_{\mathrm{eq}}=0.5$~cm, and the blue curve corresponds to the asymptotic early-time behavior (which is independent of $R_{\mathrm{eq}}$) predicted by their model.}

\dave{If we were to retain the modified edge dissipation term \cref{eq:dissipation_partial} in our own model and focus on the early stage of droplet spreading\footnote{\dave{We note that a slightly-more-involved analysis (incorporating certain elements of \citet{Durian2022}) would yield an evolution equation appropriate for characterizing the full (early and late stage) dynamics.}} ($R_\mathrm{eq}\gg R_1$), then we would find an equation of the form
\begin{equation}\label{eq:partialwetting_full}
    \left(\alpha\frac{R_1^3I_0(R_1)-2R_1^2I_1(R_1)}{V^3(I_0(R_1)-1)} + \frac{\kappa}{V^2}\right)\left(R_1^2\frac{I_0(R_1)}{I_1(R_1)} - 2R_1\right)^2\dot{R}_1 =\mathrm{const.},
\end{equation}
with the constant determined by the wetting properties of the system. It can be verified that this equation recovers the limits \cref{eq:partial_large} and \cref{eq:partial_small} in the large- and small-droplet limits, respectively. However, we note that the derivation of \cref{eq:partialwetting_full} depends on scaling arguments, just as in our model for total wetting, so should not be expected to yield exact predictions for the small-to-large-droplet transition.}


\dave{Despite these limitations, examination of the data of \citet{Cazabat1986} indicates that our model yields a close match to observations of the small-to-large transition for droplets totally wetting smooth substrates. \cref{fig:priorwork}(\textit{c}) depicts the data recorded by Cazabat \& Cohen Stuart for the spreading of small droplets of methyl-terminated poly dimethylsiloxane (PDMS; $\nu = 20$~cSt, $\sigma = 22$~mN~m$^{-1}$) on smooth, hydrophilic surfaces (microscope glass slides). As each droplet crosses the threshold between small and large horizontal radius (at a value $R_1\sim\ell_c$ that is weakly volume-dependent), Cazabat \& Cohen Stuart report a transition from $R_1\sim t^{1/10}$ to $R_1\sim t^{1/8}$. We show two of their experiments for which the transition is most visible, corresponding to drop volumes $V=0.78$~\SI{}{\micro\liter} and $V=1.5$~\SI{}{\micro\liter}. For each, we show asymptotic $R_1\sim t^{1/10}$ and $R_1\sim t^{1/8}$ scalings that approximately fit the early- and late-time data (dashed red curves). We also show our own predictions based on numerical integration of \cref{eq:hoffman_u}, augmented with all three dissipation terms present in \cref{eq:totaldissipation}, and the fitted values of $\alpha=1$, $\beta=15$, $\kappa = 150$. One can confirm that $\beta/\tan\theta\gtrsim \kappa$ for all time, as required for the small-droplet limit to exhibit $R_1\sim t^{1/10}$ growth (see~\cref{sec:contactline}).}


\section{Discussion}\label{sec:discussion}
The present work suggests a new physical picture for the spreading of viscous droplets over flat substrates, either rough or smooth. In the small droplet limit, our model \dave{converges to that of \citet{Hervet1984}. It thereby recovers the scalings $\dot{R}_\mathrm{drop}\sim \theta^3$ and $R_\mathrm{drop}\sim V^{3/10}t^{1/10}$ of \citet{Hoffman1975}, \citet{Voinov1976}, and \citet{Tanner1979} for small droplets totally wetting flat substrates}, with $V$ the droplet volume and $\theta$ its evolving contact angle. In the large droplet limit, our model provides a self-consistent, \dave{edge}-driven alternative to the predominant gravity-driven paradigm \citep{Lopez1976,Bonn2009,Popescu2012}. The resulting \emph{capillary currents} can be distinguished from traditional gravity currents through the droplet's pressure profile, which directly influences the spreading of \dave{the Darcy precursor film on a rough substrate}. The observed scaling of the precursor film radius would thus seem to support the inference of edge-driven capillary currents. 



There are various natural extensions to the present model. For one, much of the analysis for a 1-D droplet-film system follows similarly to the 2-D case. If we write $x$ for the horizontal spatial coordinate and assume the droplet to be symmetric across $x=0$, we find expressions for the droplet height profile and volume,
\[h(x) = \frac{\cosh(R_1) - \cosh(x)}{\sinh(R_1)}\,\tan\theta,\qquad V = 2(R_1\coth(R_1) - 1)\tan\theta,\]
the following analogues to \cref{eq:degennes} and \cref{eq:hoffman_u}:
\[
    R_1\coth(R_1/2)\dot{R}_1 = \alpha\theta^3_d,\qquad R_1\coth(R_1/2)\dot{R}_1 = \alpha\left(\frac{V/2}{R_1\coth R_1 - 1}\right)^{3},
\]
and the following evolution equation for the precursor film:
\[\dot{R}_2 = \frac{k}{2\phi}\,\frac{V(R_1 - \tanh R_1)^{-1} + 2s}{R_2-R_1},\]
all written in \dave{the dimensionless units \cref{eq:units}}. The right-hand side of \cref{tab:kd} reports scalings for this system. Notably, we recover the $R_1\sim V^{3/7}t^{1/7}$ scaling for thin fluid stripes reported \dave{in the experiments of \cite{Mchale1995} and first predicted by \citet{Tanner1979}}.

\dave{Our discussion in \cref{sec:priordata} also suggests how the present work might inform the modeling of \emph{partially wetting} droplets, and shows that our predictions for partially wetting droplets are consistent with those of \citet{Durian2022}. We have restricted attention here to the early spreading regime ($R_1\ll R_\mathrm{eq}$), and have reported data only for a liquid-substrate pair with very small equilibrium contact angle ($\theta_\mathrm{eq}\ll 1$). Further experiments could test the predictions of our edge-driven model for liquid-substrate pairs with larger equilibrium contact angle, or for the relaxation of a partially wetting drop into equilibrium.} \dave{Finally, a few} of the scalings predicted in \cref{tab:kd} \dave{have} yet to be observed: the $R_1\sim V^{3/5}t^{1/5}$ scaling predicted for thick fluid stripes and the $\Delta R\sim V^{-1/7}t^{2/7}$ and $\Delta R\sim V^{1/5}t^{2/5}$ scalings predicted for pressure-driven precursor films in 1-D. These predictions could conceivably be tested with droplet spreading experiments similar to our own.

\subsection*{Acknowledgments}
DD acknowledges the support of an NDSEG Graduate Fellowship. LW acknowledges the support of the Natural Sciences and Engineering Research Council of Canada (NSERC), [PDF-587339-2024]. Cette recherche a \'{e}t\'{e} financ\'{e}e par le Conseil de recherches en sciences naturelles et en g\'{e}nie du Canada (CRSNG), [PDF-587339-2024].

\subsection*{Declaration of Interests}
The authors report no conflicts of interest.

\appendix

\section{Eliminating Free Parameters in Our Model Fit}\label{sec:estimates}
We can rationalize the fitting parameters \dave{inferred from} the experiments of \cref{sec:experiment} by appropriately matching coefficients against our predictions from \cref{sec:model}. The observed scalings of $R_1$ in \cref{sec:experiment} yield the coefficient $\alpha\approx 3.45$ in the relation \cref{eq:degennes}, independent of surface roughness. \dave{Now, comparing the fitted curve $R_2(t)\sim At^{3/8}/[\log(Bt + C)]^{1/2}$ from \cref{sec:experiment} against the predicted expression \cref{eq:logthing_afterapprox} for $R_2(t)$, after tracking constants through the analysis of \cref{sec:R2-scaling} and reintroducing units, shows that 
\[A=\left(16/3\right)^{1/2}(\pi/2)^{3/8}(k/\phi)^{1/2}\alpha^{1/8}V^{1/8}\ell_c^{-3/4}u_c^{3/8}.\]
Plugging in the values $\alpha=3.45$ and $V=5$~\SI{}{\micro\liter} yields $k/\phi\approx 2.03\times 10^{-5}$~cm$^2$ for the surface S100 and $k/\phi\approx 3.28\times 10^{-5}$~cm$^2$ for S60. These estimates are consistent with known values of the permeability for similar materials ({\it e.g.} sand, fine gravel)~\citep{Bear1988}, and they satisfy the predicted ratio of grit numbers:
\[\frac{k_\mathrm{S100}}{k_\mathrm{S60}} = \frac{2.03\times 10^{-5}\;\SI{}{\centi\meter}^2}{3.28\times 10^{-5}\;\SI{}{\centi\meter}^2} = 0.6\pm 2\% = \frac{60}{100} \pm 2\%.\]
Moreover, the porosity value $k/\phi\approx 3.28\times 10^{-5}$~cm$^2$ inferred for S60 aligns closely with its measured auto-correlation length $L_\mathrm{S60}=56.7$~\SI{}{\micro\meter}, specifically $\sqrt{k_\mathrm{S60}} = 57.2~\SI{}{\micro\meter} = L_\mathrm{S60}\pm 1\%.$
Furthermore, assuming the ratio $k_\mathrm{S100}/k_\mathrm{S60}=60/100$ and inferring the value of $A$ for each surface, we observe that the predicted value for the ratio $A_\mathrm{S100}/A_\mathrm{S60}$ holds within $2\%$. Assuming $k_\mathrm{S60}=L_\mathrm{S60}^2$ and inferring the value of $A$ for S60 yields the measured value within $1\%$.} 

\dave{In turn, we can rationalize our measured value of $B$ in the fit $R_2(t)\sim At^{3/8}/[\log(Bt + C)]^{1/2}$. We neglect the time shift $C$, as its effect largely disappears for $t\gtrsim 10$~s. To estimate $B$, we revise the approximation made in \cref{eq:logthing}. Writing $\ovl{R}_1 = R_1(t=0)\approx 3.0$ mm, we now estimate $\log(R_2/R_1)\sim \log(R_2/\ovl{R}_1)$ to recover
\[R_2\sim \frac{\beta t^{3/8}}{\sqrt{\log \beta^2t^{3/4}/(\ovl{R}_1)^2}} = \frac{\sqrt{4/3}\,\beta t^{3/8}}{\sqrt{\log \beta^{8/3}(\ovl{R}_1)^{-8/3}t}}\]
for some \dave{as-of-yet unknown $\beta>0$. Setting $A = (4/3)^{1/2}\beta$}, we thus predict the value of $B$ to be  $B_\mathrm{est.}=(3/4)^{4/3}(A/\ovl{R}_1)^{8/3}$. For S100, this yields $B_\mathrm{est.}= 1.28$~s$^{-1}$, which compares favorably to the best-fit value $B=1.33$~s$^{-1}$. For S60, this yields $B_\mathrm{est.}=0.67$~s$^{-1}$, again comparable to the best-fit value $B=0.63$~s$^{-1}$.}

\bibliographystyle{jfm}
\bibliography{foo}


\end{document}